\input{aipcheck}

\documentclass[
    ,final            
  ]
  {aipproc}

\layoutstyle{6x9}

\begin{document}

\title{Electroweak symmetry breaking in TeV-scale string models}

\author{Noriaki Kitazawa}{
  address={Department of Physics, Tokyo Metropolitan University,\\
           Hachioji, Tokyo 192-0397, Japan\\
           e-mail: kitazawa@phys.metro-u.ac.jp}
}

\begin{abstract}
We propose a scenario of
 the electroweak symmetry breaking by one-loop radiative corrections
 in a class of string models with
 D3-branes at non-supersymmetric orbifold singularities
 with the string scale in TeV region.
As a test example,
 we consider a simple model based on a D3-brane
 at locally ${\bf C}^3/{\bf Z}_6$ orbifold singularity,
 and the electroweak Higgs doublet fields are identified
 with the massless bosonic modes of the open string
 on that D3-brane.
They have Yukawa couplings with three generations of
 left-handed quarks and right-handed up-type quarks
 which are identified with the massless fermionic modes
 of the open string on the D3-brane.
We calculate the one-loop correction to the Higgs mass
 due to the non-supersymmetric string spectrum and interactions,
 and qualitatively suggest that the negative mass squared
 can be generated.
The problems
 which must be solved to proceed quantitative calculations
 are pointed out.
\end{abstract}

\maketitle

\section{Introduction}

The dynamics of the electroweak symmetry breaking is still unknown.
The standard model has the naturalness or fine-tuning problem,
 and its minimal supersymmetric extension also requires
 a certain level of fine-tuning\cite{Kane:2002ap}.
Although the technicolor dynamics (without supersymmetry)
 \cite{Weinberg:1979bn,Susskind:1978ms}
 is a candidate of the ``inevitable'' electroweak symmetry breaking,
 like the chiral symmetry breaking in QCD,
 it has some problems to be overcome
 (for recent works,
  see refs.\cite{Hong:2004td,Christensen:2005cb}, for example).
The dynamical electroweak symmetry breaking
 using controllable dynamics of supersymmetric gauge theories
 is proposed in refs.\cite{Harnik:2003rs,Haba:2004bz}.
 
In this article
 we propose an alternative dynamics of
 ``inevitable'' electroweak symmetry breaking in string theory
 without supersymmetry with the string scale in TeV region.
This is the idea
 which has already proposed in ref.\cite{Antoniadis:2000tq}.
The authors calculated
 one-loop effective potential for certain scalar fields
 in a non-supersymmetric brane-anti-brane system
 ($D9$-$\overline{D5}$ system),
 and found that these scalar fields can have
 non-zero vacuum expectation values.
These scalar fields correspond to Wilson lines and brane moduli,
 and the one-loop effective potential can be obtained
 by a modification of the open string vacuum amplitude
 at one loop.
The vacuum expectation values of these scalar fields
 break original gauge symmetry USp$(16) \times$USp$(16)$.
In this letter
 we consider the similar phenomena in different systems of
 D-branes at singularities\cite{Aldazabal:2000sa},
 in which the Higgs doublet fields do not correspond to
 Wilson lines and brane moduli.
The two point function of the Higgs doublet fields,
 namely the correction to the Higgs masses, 
 have to be directly calculated to see
 whether they can have vacuum expectation values or not.

In the next section,
 we briefly review the system of D3-branes at orbifold singularities.
In the third section,
 we calculate the one-loop two point function of
 the gauge boson from the open string in ten dimensions.
This calculation
 gives a good guidance to the calculation
 of the Higgs mass correction. 
In the last section,
 we calculate the one-loop correction to
 the mass of the Higgs doublet field
 which is realized on the D3-brane at non-supersymmetric
 locally ${\bf C}^3/{\bf Z}_6$ orbifold singularity. 
We suggest that the negative mass squared can be generated.
Some concluding comments are also included in this section.

\section{D-branes at singularities}

We introduce one specific semi-realistic system
 based on a D3-brane at locally ${\bf C}^3/{\bf Z}_6$
 orbifold singularity.
A complete and self-contained introduction to
 the system of D3- and D7-branes located at singularities,
 is given in ref.\cite{Aldazabal:2000sa}.

Consider type IIB theories
 with six dimensions compactified to orbifolds or orientifolds,
 and assume that
 there is a locally ${\bf C}^3/{\bf Z}_6$ orbifold singularity
 in the compact six dimensional space.
The open string states on the D3-brane at that singularity
 are modified by the ${\bf Z}_6$ projection. 
If we simply have $N$ coincident D3-brane,
 the massless states in its four-dimensional world-volume
 belong to a ${\cal N}=4$ supersymmetric U$(N)$ gauge multiplet.
The ${\bf Z}_6$ projection
 breaks the supersymmetry and the gauge symmetry
 (to the group with the same rank,
  U$(N_1) \times$U$(n_2) \times$U$(n_3)$
  with $n_1 + n_2 + n_3 = N$, for example),
 and some states of the original gauge fields
 remain the states of gauge fields,
 some other states correspond to massless matter fields in
 bi-fundamental representations under the new gauge symmetry,
 and there are some states which are completely projected out.
If the projection keeps four-dimensional ${\cal N}=1$ supersymmetry,
 the resultant numbers of bosonic and fermionic degrees of freedom
 are equal, and the structure of ${\cal N}=1$ supermultiplet remains.
There are some consistent ways of projection
 in which there is no correspondence between
 bosonic and fermionic states
 and supersymmetry is completely broken by the spectrum.

In the language of ${\cal N}=1$ supermultiplets
 the original massless field contents on $N$ D3-brane is
 a U$(N)$ gauge vector multiplet and three chiral multiplets
 in the adjoint representation under U$(N)$.
The scalar components of three chiral multiplets
 can be understood as the position moduli of D3-brane,
 and they can be understood as
 the local complexified coordinates of
 the compact six dimensional space.
There is SU$(4)_R$ global symmetry
 under which these six bosonic degrees of freedom belong sextet
 and the four fermionic degrees of freedom,
 three chiral fermions in three chiral multiplets and gaugino,
 belong quartet. 
If the ${\bf Z}_6$ transformation is the subgroup of SU$(4)_R$,
 the projection is consistent under the time evolution.
The ${\bf Z}_6$ transformation
 on the three local complexified coordinates
 is described by three integers $b_1, b_2, b_3 = 0,1,\cdots,5$
 with the transformation matrix
 diag$(e^{2 \pi i b_1/6}, e^{2 \pi i b_2/6}, e^{2 \pi i b_3/6})$.
This is the transformation matrix on a sextet
 for any possible values of $b_i$.
The ${\bf Z}_6$ transformation
 on four fermionic degrees of freedom
 is described by four integers $a_1,a_2,a_3,a_4 = 0,1,\cdots,5$
 with $a_1+a_2+a_3+a_4=0$ mod $6$
 with the transformation matrix
 diag$(e^{2 \pi i a_1/6}, e^{2 \pi i a_2/6},
       e^{2 \pi i a_3/6}, e^{2 \pi i a_3/6})$.
Two sets of integers have the relations of
 $b_1=a_2+a_3$, $b_2=a_3+a_1$ and $b_3=a_1+a_2$.
The condition to have ${\cal N}=1$ supersymmetry is $a_4=0$,
 and in this case $b_r=-a_r$ with $r=1,2,3$. 
Both three complexified bosonic and
 three complexified fermionic open string world-sheet fields,
 which correspond to three local complexified coordinates
 in compact six dimensional space,
 transform in the same way as three local complexified coordinates.
The transformation of the space-time fermion
 is realized by the various spin combinations
 of the vacuum states of Ramond sector.

The open string Chan-Paton factor may transform under ${\bf Z}_6$.
The transformation is described as
\begin{equation}
 \vert ij \rangle \longrightarrow
  (\gamma_3)_{ii'} \vert i'j' \rangle (\gamma_3^{-1})_{j'j},
\end{equation}
 where
 $\gamma_3=\mbox{diag}
 (I_{n_0}, e^{2 \pi i / 6} I_{n_1}, \cdots
  e^{2 \pi i \cdot 5 / 6} I_{n_5})$
 with $n_0+n_1+\cdots+n_5=N$ and $I_n$ is $n \times n$ unit matrix.
 
The massless open string states,
 which are singlet under ${\bf Z}_6$ transformation,
 correspond to the massless fields.
In addition to the gauge bosons of the gauge symmetry of
 U$(n_0) \times$U$(n_1) \times \cdots \times$U$(n_5)$,
 we have massless matter fields in bi-fundamental representation:
\begin{eqnarray}
 &\mbox{complex scalars}& \qquad
  \sum_{r=1}^3 \sum_{i=0}^5 (n_i, {\bar n}_{i-b_r}),
\\
 &\mbox{Weyl fermions}& \qquad
  \sum_{\alpha=1}^3 \sum_{i=0}^5 (n_i, {\bar n}_{i+a_\alpha}),
\end{eqnarray}
 where $n_i$ and ${\bar n}_i$
 mean fundamental and anti-fundamental representation of U$(n_i)$,
 respectively.

We proceed to much more concrete model.
We take $N=6$ and $n_0=1$, $n_1=3$, $n_2=2$ and $n_3=n_4=n_5=0$.
For ${\bf Z}_6$ projection,
 we take $b_1=b_2=b_3=2$, $a_1=a_2=a_3=1$ and $a_4=-3$.
This set up gives non-supersymmetric spectrum.
We have gauge symmetry of U$(3) \times$U$(2) \times$U$(1)$
 and massless matter fields
\begin{eqnarray}
 &\mbox{Higgs doublet fields:} \quad H_r& \qquad
  3 \times (1, 2, -1),
\\
 &\mbox{left-handed quarks:} \quad q_{Lr}& \qquad
  3 \times (3, 2^*, 0),
\\
 &\mbox{right-handed quarks:} \quad u^c{}_{Lr}& \qquad
  3 \times (3^*, 1, +1),
\end{eqnarray}
 where we omit to describe the charges of U$(1)$ factors of
 U$(3)$ and U$(2)$.
There are Yukawa couplings among these fields
 which are obtained by the ${\bf Z}_6$ projection of
 the interactions due to the superpotential
 in the original ${\cal N}=4$ supersymmetric theory.
The Yukawa coupling constants
 are equal to the gauge coupling constant.
Note that
 all the gauge coupling constants of U$(3) \times$U$(2) \times$U$(1)$
 are equal at tree level.

There are four point Higgs self-couplings
 which are remnants of the D-term scalar potential
 in the original supersymmetric theory.
\begin{equation}
 V = {{g^2} \over 4}
     \sum_{r,s=1,2,3}
     \left(
      (H^\dag_r H_s)(H^\dag_s H_r)
      +
      (H^\dag_r H_r)(H^\dag_s H_s)
     \right),
\label{potential}
\end{equation}
 where $g$ is the gauge coupling constant of D3-brane.
There are no flat directions
 on the vacuum expectation values of Higgs doublet fields.
This means that
 Higgs doublet fields are not D-brane moduli.

This is not the complete system.
We have to consider the R-R (Ramond-Ramond) tadpole cancellation
 to make a consistent string theory.
R-R tadpole cancellation conditions
 contain chiral anomaly cancellation conditions.
We have to consider both twisted and untwisted tadpoles.
Twisted tadpoles can be cancelled out
 by introducing appropriate D7-branes.
Inclusion of D7-branes means introduction of additional
 gauge symmetry and massless matter fields
 which ensure the chiral anomaly cancellation
 about the gauge symmetry on D3-brane.
To consider the untwisted tadpole cancellation
 we have to specify a concrete compactification space.
Since the construction of the concrete realistic models
 is not the issue of this letter,
 and the existence of the untwisted tadpole does not affect
 our forthcoming discussions,
 we do not discuss the untwisted tadpole cancellation.

\section{One-loop two point function in ten dimensions}

We consider two point functions of gauge bosons
 in the zero momentum limit,
 namely the mass corrections to gauge bosons.
Although we know the result:
 it should vanish because of the gauge invariance,
 it is instructive to calculate the mass correction
 to the Higgs doublet fields.

Before going to the calculation in string theory,
 it is worth to mention the calculation in four dimensional
 ${\cal N}=1$ supersymmetric U(1) gauge field theory.
The quadratically divergent contributions come from
 two boson loop diagrams and one fermion loop diagram:
\begin{eqnarray}
 \Pi^{\mu\nu}_{\rm boson}
  &=&
  (1 - 2) \times
  \eta^{\mu\nu} \int {{d^4 p} \over {(2 \pi)^4 i}}
   \ {1 \over {p^2}},
\label{4D-boson}\\
 \Pi^{\mu\nu}_{\rm fermion}
  &=&
  \eta^{\mu\nu} \int {{d^4 p} \over {(2 \pi)^4 i}}
   \ {1 \over {p^2}},
\label{4D-fermion}
\end{eqnarray}
 where $\eta^{\mu\nu} = {\rm diag}(-1,1,1,1)$.
These are the contributions of one chiral multiplet.
We see the cancellation of these two corrections
 due to supersymmetry.
If we take Euclidean momentum cutoff regularization,
 boson and fermion give positive and negative correction
 to the mass squared of the gauge boson, respectively.
If we take the gauge invariant regularization,
 dimensional regularization, for example,
 the corrections vanish individually.
 
We use the techniques for one-loop calculation
 in open string theory,
 which is described in the text book of ref.\cite{GSW},
 because it has good correspondence with the one-loop
 calculation in field theory.
The one-loop contributions
 of the bosonic states and fermionic states
 to the two point function of the massless vector mode
 of the open string (gauge field) are respectively described by
\begin{eqnarray}
 A^{\rm NS} &=&
 \int {{d^{10}p} \over {(2\pi)^{10} i}}
 {\rm tr} \left( \Delta V(1) \Delta V(1) P_{\rm GSO} \right),
\label{10D-NS}\\
 A^{\rm R} &=&
 - \int {{d^{10}p} \over {(2\pi)^{10} i}}
 {\rm tr} \left( S W(1) S W(1) P_{\rm GSO} \right),
\label{10D-R}
\end{eqnarray}
 where NS and R mean Neveu-Schwarz and Ramond sector, respectively,
 $P_{\rm GSO}$ is the Gliozzi-Scherk-Olive projection operator,
 propagators are defined as
\begin{eqnarray}
 \Delta &\equiv& \int_0^1 x^{L_0 - 1} dx,
\\
 S &\equiv& i G_0 \Delta,
\end{eqnarray}
 and vertex operators are defined as
\begin{eqnarray}
 W(1) &=&
  g_O e_\mu \psi^\mu e^{i k \cdot X},
\\
 V(1) &\equiv& \{G_0, W(1)\}
\label{vertex-W}
\nonumber\\
  &=&
  {{g_O} \over \sqrt{2 \alpha'}} e_\mu
  \left( i {\dot X}^\mu + 2 \alpha' (k \cdot \psi) \psi^\mu \right)
  e^{i k \cdot X}
\label{vertex-V}
\end{eqnarray}
 with open string coupling constant $g_O$
 and polarization vector $e_\mu$.
The argument $1$ of vertex operators means
 the complex valuable $z = \exp(-i(\sigma_1+i\sigma_2))$
 with the value of the Euclidean world-sheet coordinates
 $\sigma_1=\sigma_2=0$.
Taking the value of $\sigma_1=0$ means
 that we are considering planner diagrams:
 both two vertex operators are attached to one of two boundaries
 of annulus (or cylinder).
We do not discuss the non-planner diagram,
 because it is not important to the calculation
 of the mass correction to the Higgs doublet fields
 in the next section.
The dot operation in eq.(\ref{vertex-V})
 means the differentiation by $\sigma_2$.

We evaluate
 only the leading terms in the internal momentum integration
 of eqs.(\ref{10D-NS}) and (\ref{10D-R}).
These are dominant contributions because of the following reasons.
The open string one-loop calculation is essentially
 to count the number of possible states in the loop
 with weight $\exp(-2 \pi t L_0)$
 and to integrate over the cylinder modulus $0 \le t < \infty$
 ($2 \pi t$ is the circumference of the cylinder).
The integrant of the modulus integration
 mainly has value in the regions of small $t$
 because of the exponential weight.
The leading internal momentum integration
 gives an enhancement factor for the small $t$ region,
 which is absent in the sub-leading momentum integration.
Therefore the leading term in the internal momentum integration
 can be considered as the dominant contribution.
In four dimensional field theory,
 this corresponds to evaluate only the quadratically divergent terms
 like in eqs.(\ref{4D-boson}) and (\ref{4D-fermion}).

The results of the calculation is the following.
\begin{eqnarray}
 A^{\rm NS} &\simeq&
 {1 \over 2} {{g_O^2} \over {\alpha'^5}} e^\mu e_\mu
 \int_0^1 {{d\rho} \over \rho}
 \left( {\pi \over {-\ln\rho}} \right)^5
 {1 \over {\eta(it)^8}}
 \left(
  \left( {{\theta_3(it)} \over {\eta(it)}} \right)^4
  -
  \left( {{\theta_4(it)} \over {\eta(it)}} \right)^4
 \right),
\\
 A^{\rm R} &\simeq&
 - {1 \over 2} {{g_O^2} \over {\alpha'^5}} e^\mu e_\mu
 \int_0^1 {{d\rho} \over \rho}
 \left( {\pi \over {-\ln\rho}} \right)^5
 {1 \over {\eta(it)^8}}
  \left( {{\theta_2(it)} \over {\eta(it)}} \right)^4,
\end{eqnarray}
 where $\rho = \exp(- 2 \pi t)$,
 $\eta$ is Dedekind eta function
 and $\theta_2$, $\theta_3$ and $\theta_4$
 are Jacobi theta functions.
Here we have taken the limit of zero external momentum.
The total correction vanishes because of the relation
 $(\theta_3)^4-(\theta_4)^4-(\theta_2)^4=0$,
 which means the balance of the number of
 bosonic and fermionic states, namely supersymmetry.
More rigorous calculation
 requires to include the contribution from
 non-planner and nonorientable diagrams in type I theory
 with the care of Chan-Paton factor
 as well as the sub-leading contribution
 of the internal momentum integration.
However,
 this level of calculation is enough
 to give a guide to the calculation of
 the one-loop mass correction of the Higgs doublet fields.
In case of no supersymmetry in the spectrum,
 boson and fermion contributions are not necessary balanced.
If the boson contribution is dominant,
 the correction to the mass squared is positive,
 and if the fermion contribution is dominant,
 the correction to the mass squared is negative.

\section{One-loop correction to the Higgs mass}

The calculation is very similar to that in the previous section.
The essential difference
 is the boundary condition of the world-sheet fields
 as well as ${\bf Z}_6$ projection.
Among the ten world-sheet boson and ten world-sheet fermion fields,
 the fields corresponding to the parallel direction to D3-brane
 follow Neumann boundary condition,
 and others, namely
 the fields corresponding to the transverse direction to D3-brane
 follow Dirichlet boundary condition
 for both edges of the open string.
One important fact is that
 there is no string center of mass momentum
 of the transverse direction to D3-brane.

The amplitudes to be calculated are as follows.
\begin{eqnarray}
 A^{\rm NS}_{\rm Higgs} &=&
 \int {{d^4p} \over {(2\pi)^4 i}}
 {\rm tr} \left(
           \Delta V(1)^{(-)} \Delta V(1)^{(+)} P_{\rm GSO}
            P_{{\bf Z}_6}
          \right),
\label{Higgs-NS}\\
 A^{\rm R}_{\rm Higgs} &=&
 - \int {{d^4p} \over {(2\pi)^4 i}}
 {\rm tr} \left(
           S W(1)^{(-)} S W(1)^{(+)} P_{\rm GSO}
            P_{{\bf Z}_6}
          \right),
\label{Higgs-R}
\end{eqnarray}
 where $P_{{\bf Z}_6}$ is the $Z_6$ projection operator,
 vertex operators are defined as
\begin{eqnarray}
 W(1)^{(+)} &=&
  g_O u^{i_1}{}_{i_2} \psi^{(+)} e^{i k \cdot X},
\\
 V(1)^{(+)} &\equiv& \{G_0, W(1)^{(+)}\}
\label{Higgs-vertex-W}
\nonumber\\
  &=&
  {{g_O} \over \sqrt{2 \alpha'}} u^{i_1}{}_{i_2}
  \left(
   i {\dot X}^{(+)} + 2 \alpha' (k \cdot \psi) \psi^{(+)}
  \right)
  e^{i k \cdot X}
\label{Higgs-vertex-V}
\end{eqnarray}
 with
\begin{equation}
 X^{(\pm)} = 
  {1 \over \sqrt{2}} \left( X^4 \pm i X^5 \right),
\qquad
 \psi^{(\pm)} =
  {1 \over \sqrt{2}} \left( \psi^4 \pm i \psi^5 \right)
\end{equation}
 for one of three Higgs doublet states,
 and the momentum $k_\mu$ can have non-zero value
 only for $\mu=0,1,2,3$.
The vertex operators
 $W(1)^{(-)}$ and $V(1)^{(-)}$
 are Hermite conjugates of $W(1)^{(+)}$ and $V(1)^{(+)}$,
 respectively.
The indexes $i_1$ and $i_2$ of the factor $u^{i_1}{}_{i_2}$
 denote the Chan-Paton factor of U$(1)$ and U$(2)$, respectively. 
By considering the flow of the Chan-Paton charge,
 it is easily understood that only the planner diagram contributes.

Now,
 we calculate the leading terms of internal momentum integration,
 which are dominant contributions.
We find that
 there is no leading term in Neveu-Schwarz amplitude (boson loop),
 because $X(1)^{(\pm)}$ has no zero modes
 due to Dirichlet boundary condition.
Therefore,
 we may only calculate Ramond amplitude (fermion loop),
 and make sure whether it vanishes or not.
If it is not zero,
 we have a negative correction to the Higgs mass squared,
 and the electroweak symmetry breaking
 by the Higgs vacuum expectation value
 through the one-loop quantum effect is expected.
With a special care of ${\bf Z}_6$ projection,
 we obtain the following result.
\begin{eqnarray}
 A^{\rm R}_{\rm Higgs} &\simeq&
 - 3 \cdot {1 \over 2} {{g_O^2} \over {\alpha'^2}}
   u^{\dag i_2}{}_{i_1} u^{i_1}{}_{i_2}
   \int_0^1 {{d\rho} \over \rho}
              \left( {{\pi} \over {- \ln \rho}} \right)^2
\label{mh-D3D3}
\\
&&
   \times {1 \over 6} \sum_{\gamma=0}^5
   16 \prod_{m=1}^\infty
    \left( {{1+\rho^m} \over {1-\rho^m}} \right)^2
    \left( {{1+(e^{2 \pi i / 3})^\gamma \rho^m}
             \over
             {1-(e^{2 \pi i / 3})^\gamma \rho^m}} \right)^3
    \left( {{1+(e^{- 2 \pi i / 3})^\gamma \rho^m}
             \over
             {1-(e^{- 2 \pi i / 3})^\gamma \rho^m}} \right)^3,
\nonumber
\end{eqnarray}
 where the first factor $3$ is the Chan-Paton factor.
The only $U(3)$ charged stats,
 including massless left-handed quarks
 and right-handed up-type quarks,
 contribute the loop.
Although certainly this amplitude is not zero,
 we have to consider more model dependent contributions
 due to the twisted R-R tadpole cancellations.
That is a required contribution
 which completely cancels the one-loop correction
 in case with supersymmetry.

Here, to cancel the twisted R-R tadpoles,
 we introduce 36 D7-brane
 whose world-volume is our four dimensional space-time
 and second and third complex planes
 in six dimensional compact space.
We take the ${\bf Z}_6$ action
 to the Chan-Paton factor of this D7-brane as
 $\gamma_7=\mbox{diag}
 (I_{u_0}, e^{2 \pi i / 6} I_{u_1}, \cdots
  e^{2 \pi i \cdot 5 / 6} I_{u_5})$
 with $u_0=6$, $u_1=0$, $u_2=3$ and $u_3=u_4=u_5=9$.
This D7-brane gives new gauge symmetries
 U$(6) \times$U$(3) \times$U$(9)_1 \times$U$(9)_2 \times$U$(9)_3$
 with very small gauge coupling constants,
 since we take the string scale in TeV range.
This symmetries emerge as global symmetries at low energies.
We have new massless and massive states form the open string
 with one edge on D3-brane and another edge on D7-brane.
Although we do not explain detailed spectrum here,
 since the concrete model building is not the aim of this letter,
 we would like to stress that
 there are no massless fermion states which have Yukawa couplings
 with Higgs doublet fields.
This is an important difference from the case with supersymmetry.
The leading one-loop correction to the Higgs mass squared
 from this open string is obtained as follows. 
\begin{eqnarray}
 A'^{\rm R}_{\rm Higgs} &\simeq&
 - 9 \cdot {1 \over 2} {{g_O^2} \over {\alpha'^2}}
   u^{\dag i_2}{}_{i_1} u^{i_1}{}_{i_2}
   \int_0^1 {{d\rho} \over \rho}
              \left( {{\pi} \over {- \ln \rho}} \right)^2
   \cdot {1 \over 6} \sum_{\gamma=0}^5
   \left( (e^{2 \pi i / 3})^\gamma
        + (e^{- 2 \pi i / 3})^\gamma \right)
\nonumber\\
\qquad
&&
   \times
   16 \prod_{m=1}^\infty
    \left( {{1+\rho^m} \over {1-\rho^m}} \right)^2
    \left( {{1+(e^{2 \pi i / 3})^\gamma \rho^m}
             \over
             {1-(e^{2 \pi i / 3})^\gamma \rho^m}} \right)
    \left( {{1+(e^{- 2 \pi i / 3})^\gamma \rho^m}
             \over
             {1-(e^{- 2 \pi i / 3})^\gamma \rho^m}} \right)
\nonumber\\
\qquad
&&
   \times
    \left( {{1+(e^{2 \pi i / 3})^\gamma \rho^{m-1/2}}
             \over
             {1-(e^{2 \pi i / 3})^\gamma \rho^{m-1/2}}} \right)^2
    \left( {{1+(e^{- 2 \pi i / 3})^\gamma \rho^{m-1/2}}
             \over
             {1-(e^{- 2 \pi i / 3})^\gamma \rho^{m-1/2}}} \right)^2,
\label{mh-D3D7}
\end{eqnarray}
 where the first factor $9$ is the Chan-Paton factor,
 and the factor 
 $((e^{2 \pi i / 3})^\gamma + (e^{- 2 \pi i / 3})^\gamma)$
 means the non-trivial transformation of Ramond sector vacuum
 under the ${\bf Z}_6$ action,
 which is related with the fact that
 no massless fermion states have Yukawa couplings
 with the Higgs doublet.
Only the massive states with U$(9)_1$ or U$(9)_3$ charges contribute.
The last two factors in the infinite product
 are the realizations of the Dirichlet-Neumann boundary condition
 of the open string in second and third complex planes
 in compact six dimensional space.
The contribution of eq.(\ref{mh-D3D7}) certainly
 does not cancel out the contribution of eq.(\ref{mh-D3D3}),
 though the divergent contributions from
 twisted R-R tadpoles are cancelled out.
The finite contribution of eq.(\ref{mh-D3D7}) should be negative,
 since it is also due to the fermion one loop.
We suggest that
 the negative mass squared of the Higgs doublet field
 can be generated,
 and the electroweak symmetry breaking can be expected,
 in this class of non-supersymmetric models
 with D-branes at singularities.
 
There are some concluding comments in order.

There may still exist a divergence in the total correction of
 eq.(\ref{mh-D3D3}) plus eq.(\ref{mh-D3D7})
 due to the uncanceled twisted NS-NS 
 (Neveu-Schwarz-Neveu-Schwarz) tadpoles.
Since there is no supersymmetry,
 R-R tadpole cancellation
 does not necessary result NS-NS tadpole cancellation.
The existence of NS-NS tadpole means that
 some redefinitions of backgrounds are required
 \cite{Fischler:1986ci,Fischler:1986tb,Das:1986dy,Dudas:2004nd}. 
It is probable that
 twisted moduli obtain vacuum expectation values
 as a result of the background redefinition
 by the uncanceled twisted NS-NS tadpoles.
The vacuum expectation values of twisted moduli
 give some modification of Higgs potential of eq.(\ref{potential})
 due to the emergence of the Fayet-Iliopoulos terms
 for anomalous U$(1)$ gauge symmetries
 in the original supersymmetric theory
 \cite{Douglas:1996sw,Douglas:1997de}.
This effect by itself can give vacuum expectation value
 to Higgs doublet fields even without negative mass squared
 at one loop.
The vacuum expectation values of twisted moduli
 may also result blowing-up the orbifold singularity
 (see for example \cite{Cvetic:1999qx}),
 and the actual geometrical D3-brane reconfiguration
 by the vacuum expectation value of Higgs doublet fields
 may be understood at this blown-up orbifold singularity.

There is another possibility that the present
 non-supersymmetric ${\bf C}^3/{\bf Z}_6$ orbifold singularity
 is unstable and decays to some non-singular point.
In other words,
 there might be tachyon modes in twisted NS-NS closed string sector
 localized at the non-supersymmetric singularity,
 which means the instability of the singularity\cite{Adams:2001sv}.
The non-supersymmetric orbifold singularities
 suffer from this phenomenon in general.
We need to find
 stable non-supersymmetric singularities without tachyons
 or non-supersymmetric singularities with very long life time.

There are three Higgs doublet fields,
 $H_1$, $H_2$ and $H_3$, in our ${\bf Z}_6$ model,
 and our calculations concern one of them, $H_1$.
The contribution of eq.(\ref{mh-D3D3})
 is applicable for all three Higgs doublet fields,
 but the contribution,
 which is related with the R-R tadpole cancellations,
 may be different depending on models.
In our model with one D7-brane,
 the result of eq.(\ref{mh-D3D7}) is only for $H_1$,
 and $H_2$ and $H_3$ will obtain a different result.
This kind of asymmetric configuration of D7-branes
 for twisted R-R tadpole cancellation
 gives asymmetric corrections to Higgs doublet fields.
It may be possible that only one Higgs doublet field,
 which is a linear combination of Higgs doublet fields,
 has vacuum expectation value,
 and others are heavy.
This is a phenomenologically preferable situation,
 since the existence of many Higgs doublet fields
 with vacuum expectation values causes a problem,
 flavor-changing neutral current problem, in general.
Such a Higgs doublet field may have non-trivial Yukawa couplings
 for the hierarchical masses and flavor mixings,
 though the original Higgs doublet fields
 usually have trivial Yukawa couplings.

Inclusion of sub-leading term is important
 for more quantitative and detailed analysis.
Modern techniques,
 like path integral formalism, might be better for this aim.

\section*{Acknowledgment}

I would like to thank O.~Yasuda for useful comments.

\end{document}